\documentclass[aip,apl,reprint]{revtex4-1}

\usepackage{hyperref}
\usepackage{graphicx}

\usepackage{times}

\begin{document}

\title{Low-crosstalk bifurcation detectors for coupled flux qubits}

\author{P.~C.~de~Groot}
\email{p.c.degroot@tudelft.nl}
\affiliation{Kavli Institute of Nanoscience, Delft University of Technology, P.O. Box 5046, 2600 GA Delft, The Netherlands}

\author{A.~F.~van~Loo}
\affiliation{Kavli Institute of Nanoscience, Delft University of Technology, P.O. Box 5046, 2600 GA Delft, The Netherlands}

\author{J.~Lisenfeld}
\affiliation{Kavli Institute of Nanoscience, Delft University of Technology, P.O. Box 5046, 2600 GA Delft, The Netherlands}
\affiliation{Physikalisches Institut and DFG Center for Functional Nanostructures (CFN), Karlsruhe Institute of Technology, D-76131 Karlsruhe, Germany}

\author{R.~N.~Schouten}
\affiliation{Kavli Institute of Nanoscience, Delft University of Technology, P.O. Box 5046, 2600 GA Delft, The Netherlands}

\author{A. Lupa\c{s}cu}
\affiliation{Institute for Quantum Computing, University of Waterloo, N2L 5G7 Waterloo, Canada}

\author{C.~J.~P.~M~Harmans}
\affiliation{Kavli Institute of Nanoscience, Delft University of Technology, P.O. Box 5046, 2600 GA Delft, The Netherlands}

\author{J.~E.~Mooij}
\affiliation{Kavli Institute of Nanoscience, Delft University of Technology, P.O. Box 5046, 2600 GA Delft, The Netherlands}

\date{\today}

\begin{abstract}

  We present experimental results on the crosstalk between two AC-operated dispersive bifurcation detectors, implemented in a circuit for high-fidelity readout of two strongly coupled flux qubits. Both phase-dependent and phase-independent contributions to the crosstalk are analyzed. For proper tuning of the phase the measured crosstalk is 0.1~\% and the correlation between the measurement outcomes is less than 0.05~\%. These results show that bifurcative readout provides a reliable and generic approach for multi-partite correlation experiments.

\end{abstract}

\maketitle
\newpage

Dispersive bifurcation detectors \cite{siddiqi:04} have proven to be well suited for single-qubit state detection in superconducting devices \cite{clarke:08}. High fidelity, low back-action, and quantum nondemolition readout have been reported \cite{siddiqi:06, lupascu:07, mallet:09}. An expansion to multiple qubits and detectors is the obvious next step. This introduces additional requirements, such as sufficiently low direct detector-detector interaction to avoid measurement crosstalk \cite{mcdermott:05}. Low crosstalk is important for all experiments that rely on measurement correlations, such as detecting entanglement.

We quantitatively analyze the crosstalk between two simultaneously operated bifurcative flux detectors, each attached to a flux qubit \cite{mooij:99, qubits}. Concentrating on the detectors, the qubits are biased such that their role can be disregarded. The detectors are designed for high fidelity single-shot readout, with similar design parameters as in Ref.~\cite{lupascu:07}. We measure a very small crosstalk of less than \( 0.4~\% \), in spite of the close proximity of all components as required for strong inductive coupling between the two qubits (\( 250~\mathrm{MHz} \)) and between the qubits and their detectors (\( 90~\% \) qubit-state separability).

\begin{figure}[h!b]
  \begin{center}
    \includegraphics{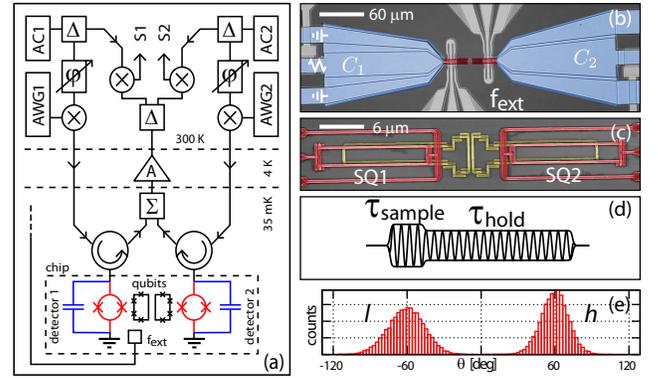}
  \end{center}
  \caption{(Color) (a) Schematic representation of the readout circuit. (b) Optical microscope picture showing the on-chip capacitors (\( C_1 = 31~\mathrm{pF} \), \( C_2 = 40~\mathrm{pF} \)) and two antennas for external driving. (c) Scanning electron micrograph showing the two SQUID loops. These loops are \( 10.7 \times 2.0~\mathrm{\mu m^2} \) with a mutual inductance of \( M_{sq-sq} = 5~\mathrm{fH} \). Each SQUID covers partially one of the two coupled qubits, with \( M_{sq-qb} = 10~\mathrm{pH} \) and \( M_{qb-qb} = 0.76~\mathrm{pH} \). The resonator loops have a mutual inductance of \( M_{res,1-res,2} = 1~\mathrm{fH} \). The couplings of the resonator loops to the SQUID and qubit loops are minimized by the symmetry of the design. The maximum critical current of each SQUID is \( I_{c,max} \approx 2.0~\mathrm{\mu A} \). The experiment is performed at the flux operating point \( \Phi/\Phi_0 = 2.67 \) . (d) Shape of the AC readout burst. (e) The histogram shows the statistics of the phase \( \theta \) picked up at the reflection of 32768 individual measurements. The two oscillator states can clearly be distinguished.}
  \label{fig:setup}
\end{figure}

  When a detector is driven, it can assume two classical states. The detector starts in one state and will or will not switch to the other state, depending on the sensed magnetic flux. We distinguish two types of crosstalk. If the \emph{driving} signal for one detector directly couples to the second one, the switching probability of that detector may be affected. We call this input driving crosstalk (IDC). In addition, the switching of one detector may be influenced by the \emph{state} of the other, leading to undesired correlation of the output states. We call this detector state crosstalk (DSC).

  Each detector comprises a superconducting quantum interference device (SQUID) as the flux-sensor and a shunting capacitor (Fig.~1(a-c)).
  The flux sensed by the SQUID controls its inductance, which, combined with the capacitor, results in a resonant circuit with a flux-dependent resonance frequency. In addition, the inductance depends on the current in the SQUID, making the resonator non-linear. When it is subjected to a strong AC driving slightly below its resonance frequency, the response shows two stable states (\( l \) and \( h \)) differing by their amplitude and phase. This type of detector is often referred to as a Josephson bifurcation amplifier \cite{siddiqi:04}. The capacitors of the two detectors are chosen such that their resonance frequencies differ by \( \sim 10~\% \).

  Each detector is operated in the same way as in a single qubit experiment. The driving signals are generated using AC-sources AC1 and AC2  and shaped into two-level bursts using mixers and arbitrary waveform generators AWG1 and AWG2 (see Fig.~1(a,d)). The first part of the burst of duration \( \tau_{sample} \) is where the actual qubit state measurement would take place. The driving amplitude is such that the resonator has a significant probability to switch from its initial low amplitude state \( l \) to the high amplitude state \( h \). The switching probability from \( l \) to \( h \) strongly depends on the resonance frequency, and therefore on the flux sensed by the SQUID loop.
  The second part of the burst has a reduced amplitude and a duration \( \tau_{hold} \). At this driving amplitude the probabilities for switching from \( l \) to \( h \) and from \( h \) to \( l \) are negligible. The resonator state is effectively frozen, which provides the required time to integrate the signal for the desired signal-to-noise ratio.
  A pulse shape was used with \( \tau_{sample} = 20~\mathrm{ns} \) and \( \tau_{hold} = 300~\mathrm{ns} \).

  The applied AC burst is fully reflected at the resonator. In the reflection, a phase is picked up that depends on the state of the resonator; this phase is measured to distinguish the resonator states.
  Before arriving at the resonator, the burst has passed several stages of attenuation for thermalization (not shown), a band-pass filter (not shown), and a circulator.
  The reflected signals of both detectors pass again through the circulators, are combined, amplified \cite{amplifier}, split up again at room temperature, and down-mixed with the original AC-signals.
  The multiplexing of the output signal significantly reduces the heat load and space required in the cryogenic setup.
  Since the down-mixed signals S1 and S2 contain multiple frequencies of both detectors, \( 30 ~\mathrm{MHz} \) low-pass filters are used to extract the desired DC components.
  The level of this signal, which is a measure of the phase picked up in the reflection, is recorded with a 100 MHz, 8-bit digitizer. The phase shifters (Fig.~1(a)) are adjusted to obtain the maximum discrimination of the two oscillator states in the down-mixed signal.
  Fig.~1(e) shows the result of 32768 measurements of the state of one of the oscillators for 50~\% switching probability; it clearly demonstrates the good separation of the two associated phases.

\begin{figure}[b]
  \begin{center}
    \includegraphics{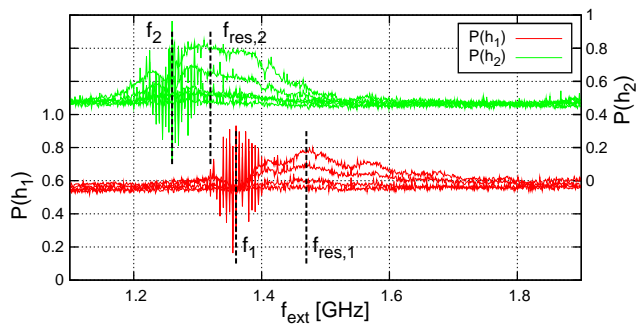}
  \end{center}
  \caption{(Color online) Switching probability for detector 1 and 2 as a function of the driving frequency of the externally applied field. Individual curves differ \( 4~\mathrm{dB} \) in driving power. The dashed lines indicate the driving frequency \( f_i \) and resonance frequency \( f_{\mathrm{res},i} \) for each detector.}
  \label{fig:mwline}
\end{figure}

  Before we proceed to quantify the crosstalk between the detectors, we discuss an experiment that qualitatively shows the effect of an external oscillating field on the detectors. The resonators are continuously driven by a nearby antenna (\( f_{ext} \) in Fig.~1(a,b)) that generates a field in the detectors that has at least 100x higher amplitude than is expected from the detector-detector crosstalk. The frequency \( f_{ext} \) is varied; for each frequency the probability \( P(h) \) of switching to state \( h \) is determined from 4096 measurements. The result is shown in Fig.~2. From this graph two types of response can be distinguished.
  The first shows up as a broad peak at the resonance frequencies \( f_{\mathrm{res},1} = 1.47~\mathrm{GHz} \) and \( f_{\mathrm{res},2} = 1.32~\mathrm{GHz} \) of the resonators. We attribute this effect to excitation of the resonator, which is consistent with the increased probability for switching to state \( h \).
  The second effect can be seen in the regions around the detector driving frequencies \( f_1 = 1.36~\mathrm{GHz} \) and \( f_2 = 1.26~\mathrm{GHz} \). In a region of approximately \( 100~\mathrm{MHz} \) wide, and when \( f_{ext} \) is an exact multiple of the measurement repetition frequency \( f_{rep} = 1/\tau_{rep} = 1/22.2~\mathrm{\mu s}^{-1} \), the probability of switching varies drastically, ranging almost from 0 to 1. This effect is attributed to interference of the detector driving current with the externally applied field from the antenna. If the frequency difference \( |f_i - f_{ext}| \) is small compared to \( 1/\tau_{sample} = 50~\mathrm{MHz} \) they remain effectively phase correlated during each single measurement and the contributions add up, enhancing or reducing \( P(h) \) depending on their phase difference. Note that qubit operation pulses would not induce this interference, as they are well separated in time and frequency from the readout pulses.
  These results clearly demonstrate effects of an AC-field, including the role of the phase. With this in mind we proceed to quantify the detector-detector crosstalk.

\begin{figure}[bt]
  \begin{center}
    \includegraphics{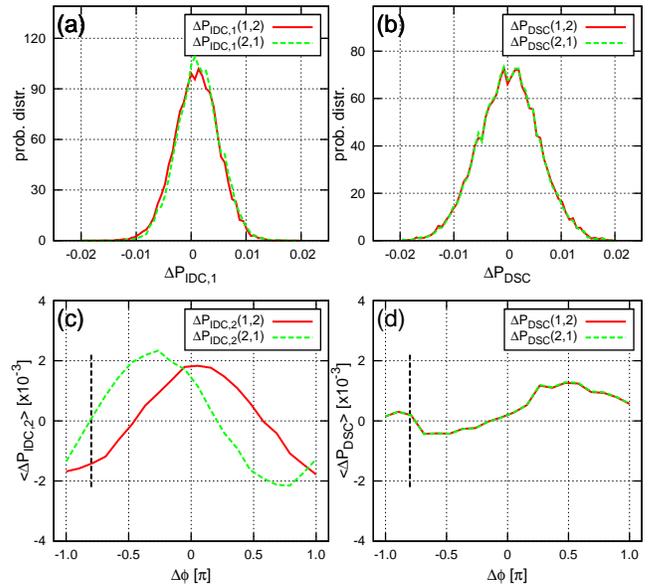}
  \end{center}
  \caption{(Color online) (a) The probability distribution for the measured Input Driving Crosstalk. The crosstalk shows as a shift of the center of the histograms: the average switching probability is changed by \( 0.10~\% \) and \( 0.13~\% \) for detector 1 and 2, respectively, if the other detector is operated. (b) The probability distribution for the Detector State Crosstalk. The average switching probability differs by \( 0.07~\% \) and \( 0.08~\% \) for detector 1 and 2, respectively, for the two different outcomes of the other detector. (c) The IDC as a function of the phase difference \( \Delta\phi \) between the two AC-driving-signals. (d) The DSC as a function of \( \Delta\phi \). The dashed vertical lines indicate the phase setting with favorable crosstalk for the total IDC and the DSC combined.}
  \label{fig:cross}
\end{figure}

  We first investigate the phase-independent components of the crosstalk. The probability \( P(h_i) \) that detector \( i \) switches to state \( h \) is determined two times, each from 32768 individual measurements. The first time detector \( j \) is not driven (\( \mathrm{OFF}_j \)) and the second time it is driven and read out with a regular readout pulse (\( \mathrm{ON}_j \)), providing also \( P(h_j) \). This sequence is repeated 10000 times.
  Since the detectors are operated at different frequencies, we define the relevant phase \( \Delta\phi \) as the phase-difference between the two AC-signals at the start of each measurement. We choose the driving frequencies \( f_1 = 1.37009999~\mathrm{GHz} \) and \(f_2 = 1.26~\mathrm{GHz} \) and the repetition time \( \tau_{rep} = 22.2~\mathrm{\mu s} \). This leads to a uniform distribution of \( \Delta\phi \) over the 32768 measurements. In a separate experiment, it was verified that for a single detector the phase of the driving-signal does not influence the measurement outcome.

  We determine the IDC, quantified as:
      \begin{equation}
        \Delta P_{\mathrm{IDC},1}(i,j) =
        P( h_i ; \mathrm{ON}_j ) - P( h_i ; \mathrm{OFF}_j )
        \label{DPIDC1}
      \end{equation}
  where \( P( h_i ; \mathrm{ON}_j ) \) and \( P( h_i ; \mathrm{OFF}_j ) \) are the switching probabilities of detector \( i \) for the cases that detector \( j \) is \( \mathrm{ON} \) and \( \mathrm{OFF} \) respectively.
  The results are shown in Fig.~3(a). We find very small crosstalk, \( \left< \Delta P_{\mathrm{IDC},1}(1,2) \right> = 0.10~\% \) and \( \left< \Delta P_{\mathrm{IDC},1}(2,1) \right> = 0.13~\% \); \( \left< \right> \) denotes the average over the 10000 repetitions.
  The IDC from detector one to two is slightly higher then vice-versa; this is consistent with the picture that the crosstalk causes an additional excitation and \( |f_1 - f_{\mathrm{res},2}| < |f_2 - f_{\mathrm{res},1}| \).
  A separate analysis of the data showed no significant increase of the width of the distribution for detector \( i \) if detector \( j \) is ON.

  The DSC is quantified as:
      \begin{equation}
        \Delta P_{\mathrm{DSC}}(i,j) =
        P( h_i | l_j ) - P( h_i | h_j )
        \label{DPDSC}
      \end{equation}
  where \( P( h_i | l_j) \) and \( P(h_i | h_j) \) are the conditional probabilities for finding detector \( i \) in the high state given that detector \( j \) was in the low or high state respectively. For this we only use the data where both detectors were ON. The results are shown in Fig.~3(b). We find \( \left< \Delta P_{\mathrm{DSC}}(1,2) \right> = 0.07~\% \) and \( \left< \Delta P_{\mathrm{DSC}}(2,1) \right> = 0.08~\% \). Again the crosstalk is very small. In both cases the DSC is positive, implying that the switching probability is decreased if the other detector switches.

  In the last experiment we measure the average switching probability as a function of \( \Delta \phi \). Now we choose the driving frequencies \( f_1 = 1.37~\mathrm{GHz} \) and \(f_2 = 1.26~\mathrm{GHz} \) to be multiples of the repetition frequency \( f_{rep} = 1/\tau_{rep} = 1/22.2~\mathrm{\mu s}^{-1} \), so that \( \Delta\phi \) is constant within a sequence of measurements. We change \( \Delta \phi \) from \( -\pi \) to \( \pi \) in 20 steps by changing the phase of AC2; this is repeated 5000 times. The switching probability is determined from 16384 individual measurements.
  To ensure phase stability of the phase-locked AC-sources during the time of the experiment (\( \sim 48~\mathrm{hours} \)), the phases are monitored by sampling the driving signals. The phases are checked after each phase-change giving a maximum phase-error of \( \pm 5~\mathrm{degrees} \). The trigger-jitter of the AWG that generates the envelope is \( 50~\mathrm{ps} \) (RMS).
   Because in this experiment both detectors are always ON, we now redefine the IDC relative to its average value:
  \begin{equation}
    \Delta P_{\mathrm{IDC},2}(i,j,\Delta\phi) =
    P( h_i ; \Delta\phi ) - \left< P( h_i ; \Delta\phi ) \right>_{\Delta\phi}
    \label{DPIDC2}
  \end{equation}
  where \( \left< \right>_{\Delta\phi} \) denotes the average over all phase-differences.

  The result for the IDC (Fig.~3(c)) shows a clear phase dependence for both detectors with an amplitude of 0.18~\% for detector 1 and 0.22~\% for detector 2. A prominent feature is the shift in \( \Delta\phi \) of \( \sim 0.3\pi \) between the two curves. This is attributed to the phase difference between the applied driving field and the internal current response of each resonator, which depends on the detuning in the employed off-resonant driving.
  The total IDC is obtained by adding \( \left< \Delta P_{\mathrm{IDC},1} \right> \) and \( \left< \Delta P_{\mathrm{IDC},2} \right> \); it remains below \( 0.4~\% \) for both detectors and has an optimum \( <0.1\% \) indicated by the dashed line in figure Fig.~3(c)
  The DSC (Fig.~3(d)) also shows a phase dependence, although less distinct, with an amplitude of 0.08~\% for both detectors. Here the phase dependence can be understood from the fact that a change of the resonator state changes its internal current, and thereby the field it induces in the other resonator. Depending on the phases, the change will enhance or reduce the switching of the other detector. For the DSC we do not find a shift in \( \Delta\phi \) between the two detectors, which is in agreement with direct resonator-resonator interaction and no contribution of the driving signal. We observe a slight positive offset of 0.04~\%, which is consistent in sign, but slightly smaller than found in the phase-independent experiment of Fig.~3(b).

  In conclusion, we have demonstrated that two bifurcative flux detectors strongly coupled to two interacting qubits can be implemented with very small crosstalk. The crosstalk depends on the phase relation between the AC-driving signals, with a maximum of 0.4~\% and an optimum of less than 0.1~\%. The results indicate that this approach likely can be extended to five detectors per octave amplifier bandwidth. Furthermore, the smaller size of lumped-element resonators compared to transmission-line resonators allows a compact single-chip multiple-device design. Adding moderate-Q resonators at the input side could provide multiplexing of the detector driving as well, allowing the use of one single coaxial feed line for all detectors.

  The authors thank I. T. Vink and the Delft Flux Qubit Team for help and discussions and A. van der Enden, R. G. Roeleveld, B. P. van Oossanen and the Nanofacility for technical and fabrication support. This work is supported by NanoNed, FOM and EU projects EuroSQIP and CORNER.



\begin{thebibliography}{1}
\bibitem{qubits}
    Qubits 1 and 2 are characterized by their persistent current \( I_{p,1}=540~\mathrm{nA} \) and \( I_{p,2}=470~\mathrm{nA} \) and energy gap \( \Delta_1/h=2.1~\mathrm{GHz} \) and \( \Delta_2/h=4.8~\mathrm{GHz} \). They are biased at level splittings of \( \nu_1=7.3~\mathrm{GHz} \) and \( \nu_2=11.1~\mathrm{GHz} \) respectively.
\end{thebibliography}

\end{document}